\documentclass[a4paper,12pt]{article}

\usepackage{amstext,amsmath,amssymb,amsthm}  

\usepackage[normalem]{ulem}  
\usepackage{graphicx}  
\usepackage{bm} 

\usepackage[comma,sort,authoryear]{natbib}
\usepackage{hypernat}  

\usepackage{vmargin}  
\usepackage{authblk}

\usepackage{xcolor}
\definecolor{dark-red}{rgb}{0.8,0.15,0.15}
\definecolor{dark-blue}{rgb}{0.15,0.15,0.6}
\definecolor{medium-blue}{rgb}{0,0,0.8}

\usepackage[pdftex,breaklinks,colorlinks]{hyperref}
\hypersetup{  
  pdftitle =
    {Reducing the standard deviation in multiple-assay experiments where the variation matters but the absolute value does not},
  linkcolor={dark-red},
  citecolor={dark-blue},
  urlcolor={medium-blue}
}

\begin{document}

\numberwithin{equation}{section}
\numberwithin{figure}{section}
\allowdisplaybreaks[1]  

\title{Reducing the standard deviation in multiple-assay experiments where the variation matters but the absolute value does not}

\author[,1,2,3,4,5]{Pablo Echenique-Robba\footnote{{\footnotesize \
\href{mailto:pablo.echenique.robba@gmail.com}{\texttt{pablo.echenique.robba@gmail.com}} ---
\href{http://www.pabloecheniquerobba.com}{\texttt{http://www.pabloecheniquerobba.com}}}}}
\author[6,2,7]{Mar\'{\i}a Alejandra Nelo-Baz\'an}
\author[7,2,5]{Jos\'e A. Carrodeguas}

\affil[1]{Instituto de Qu\'{\i}mica F\'{\i}sica Rocasolano, CSIC, Madrid, Spain}
\affil[2]{Instituto de Biocomputaci\'on y F{\'{\i}}sica de Sistemas Complejos (BIFI), Universidad de Zaragoza, Spain}
\affil[3]{Zaragoza Scientific Center for Advanced Modeling (ZCAM), Universidad de Zaragoza, Spain}
\affil[4]{Departamento de F{\'{\i}}sica Te\'orica, Universidad de Zaragoza, Spain}
\affil[5]{Unidad Asociada IQFR-BIFI, Madrid-Zaragoza, Spain}
\affil[6]{Fundaci\'on Gran Mariscal de Ayacucho (Fundayacucho), La Urbina, Venezuela}
\affil[7]{Departamento de Bioqu\'{\i}mica y Biolog\'{\i}a Molecular y Celular, Universidad de Zaragoza, Spain}

\date{\today}

\maketitle

\begin{abstract}

You measure the value of a quantity $x$ for a number of systems (cells,
molecules, people, chunks of metal, DNA vectors, etc.). You repeat the whole
set of measures in different occasions or \emph{assays}, which you try to
design as equal to one another as possible. Despite the effort, you find that
the results are too different from one assay to another. As a consequence,
some systems' averages present standard deviations that are too large to
render the results statistically significant. In this work, we present a novel
correction method of very low mathematical and numerical complexity that can
reduce the standard deviation in your results and increase their statistical
significance as long as two conditions are met: inter-system variations of $x$
matter to you but its absolute value does not, and the different assays
display a similar tendency in the values of $x$; in other words, the results
corresponding to different assays present high linear correlation. We
demonstrate the improvement that this method brings about on a real cell
biology experiment, but the method can be applied to any problem that conforms
to the described structure and requirements, in any quantitative scientific
field that has to deal with data subject to uncertainty.
\vspace{0.4cm}\\ {\bf Keywords:} multiplicative systematic error, reducing standard deviation, multiple assays, inter-system variation, linear correlation, statistical significance
\vspace{0.2cm}\\

\end{abstract}

\newpage

\section{Introduction}
\label{sec:introduction}

Imagine you measure in the laboratory a given quantity $x$ for six different
\emph{systems}: system 1, system 2, \ldots, and system 6 (they could be cell
types, people, proteins or DNA vectors, even the same system at different
times if the quantity $x$ is expected to evolve in some reproducible manner).
You want to be sure that you are making no mistakes, so you repeat the whole
set of six measures three times, say, in different days (you try hard so that
the only thing that changes from one time to the next is the day). We will
call each one of these repeated experiments an \emph{assay}, in this case,
assay 1, assay 2 and assay 3. At the end of the process, you are in possession
of $6 \times 3$ values of the quantity $x$; six for each assay, three for each
system.

\begin{figure}[!ht]
\begin{center}
\includegraphics[scale=0.3]{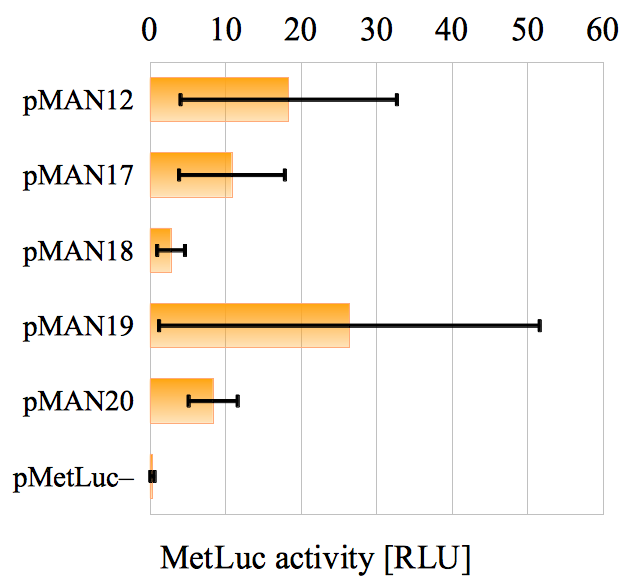}
\caption{\label{fig:errorbars_uncorrected}{\small Bar chart representation of 
the average values $\mu$ (orange bars) and the associated standard deviation
$\sigma$ (black capped lines) in tab.~\ref{tab:uncorrected_raw}.}}
\end{center}
\end{figure}

Now imagine you obtain the values in tab.~\ref{tab:uncorrected_raw} (the
strange names for the six systems in the first column will be explained
later). The first thing we can say about the results is that they do not look
good at all. The standard deviation from the average is comparable to the
average itself for most of the systems, and only on a couple of them you are
`lucky' enough so that the former is about half the value of the latter. You
check the corresponding chart in fig.~\ref{fig:errorbars_uncorrected}, and you 
see the same despairing situation. The error bars are humongous!

\begin{table}[!ht]
\begin{center}
\begin{tabular}{l||rrr|r@{\hspace{3pt}}c@{\hspace{3pt}}r}
           & assay 1 & assay 2 & assay 3 & $\mu$ & $\pm$ & $\sigma$ \\
\hline
pMAN12     &   33.88 &    5.65 &   15.53 & 18.36 & $\pm$ &    14.33 \\
pMAN17     &   17.60 &    3.61 &   11.29 & 10.83 & $\pm$ &     7.01 \\
pMAN18     &    4.62 &    0.94 &    2.72 &  2.76 & $\pm$ &     1.84 \\
pMAN19     &   55.35 &    9.30 &   14.52 & 26.39 & $\pm$ &    25.22 \\
pMAN20     &   11.15 &    4.78 &    9.10 &  8.35 & $\pm$ &     3.52 \\
pMetLuc$-$ &    0.00 &    0.39 &    0.54 &  0.31 & $\pm$ &     0.28
\end{tabular}
\caption{\label{tab:uncorrected_raw}{\small Activity of the MetLuc protein
($x$ quantity) under the control of six different promoter sequences (the six
systems) measured in three assays. The last two columns correspond to the
average $\mu$ of the three assays for each system, and the associated standard
deviation (or error) $\sigma$. The units as well as the rest of the
experiment's details are described in sec.~\ref{subsec:experiment}.}}
\end{center}
\end{table}

\begin{figure}[!ht]
\begin{center}
\includegraphics[scale=0.25]{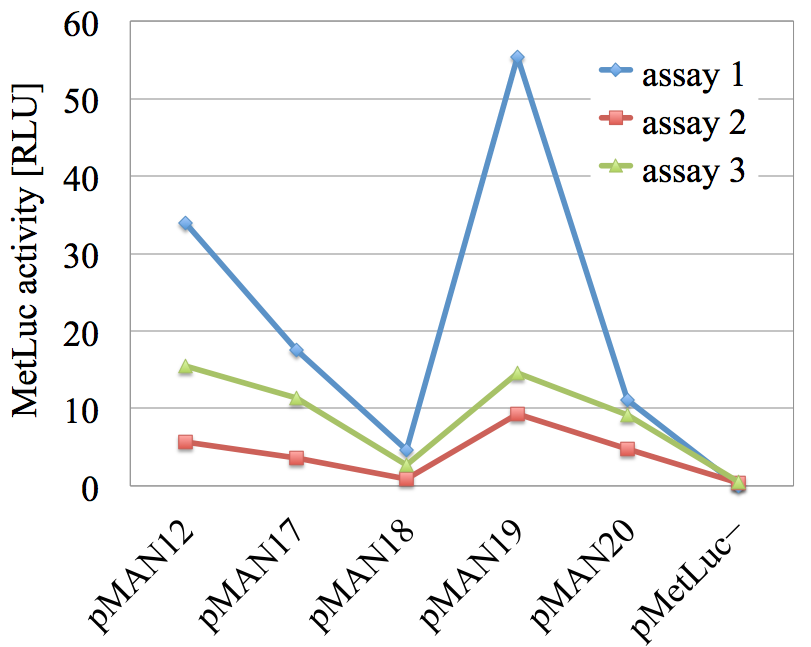}
\caption{\label{fig:tendency_uncorrected}{\small Variation of the quantity $x$ 
(MetLuc activity) in tab.~\ref{tab:uncorrected_raw} for the six systems
(vectors) studied. Each color corresponds to a different assay, and the lines 
joining the experimental points have been added for visual comfort.}}
\end{center}
\end{figure}

Before throwing in the towel, you realize two characteristics about your
experiments that might save your day:

\begin{itemize}

\item The fact is that the \emph{absolute value} of $x$ for each given system
is not really very important to you. What you are really interested in
properly measuring is the \emph{variation} in $x$ from one system to another.
For example, whether or not you could safely claim that the value of $x$
corresponding to system 1 is larger than, and approximately the double of,
that associated to system 5.

\item Even if you seem to be measuring huge differences in absolute value
across the different assays, it looks as if the `tendency' of the variations
is similarly captured in all three of them. This is even more apparent in the
graphical representation in fig.~\ref{fig:tendency_uncorrected}.

\end{itemize}

In this work, we will argue that you are right if you do not throw in the
towel in such a circumstance. We will interpret the structure of the results
as being caused by a \emph{multiplicative systematic error} (across the
different assays), and we will propose a method to \emph{correct} your results
in a way such that this systematic error is removed. As a consequence, the
corrected numbers will not tell you anything significant about the `true'
absolute value of $x$ for the different systems, but, in exchange, they will
maximally capture the tendency that you seemed to be correctly measuring. That
is, the averages of the corrected results will present appreciable smaller
standard deviations while still following the average tendency of variation.

In the next section, we will make this precise by introducing the general
method of correction as well as a real experiment in cell biology which
suffers from the problems (and the virtues!) that we have mentioned in this
introduction. In sec.~\ref{sec:results}, we will apply the correction method
to this experiment to show that both the standard deviations and the
statistical significance of the results improves considerably. In
sec.~\ref{sec:discussion}, we will discuss our interpretation of the studied
situation and the proposed method, we will compare it to a simpler
alternative, and we will try to explain the surprising fact that something so
straightforward cannot be found (as far as we are aware) in the previous
literature. Finally, in sec.~\ref{sec:conclusions}, we will briefly summarize
the main conclusions of this work, and we will outline some open questions and
lines of future research

\section{The method and a real example}
\label{sec:method}

\subsection{Experimental setup}
\label{subsec:setup}

As we advanced we will have in general $N$ systems, among which a specific one
may be called \emph{system} $j$, with $j=1,2,\ldots,N$. We now measure a
quantity $x$ for each one of the $N$ systems, and we repeat $M$ times the
whole set of $N$ measures. A generic repetition is termed \emph{assay} $k$,
with $k=1,2,\ldots,M$, and each one of them is carried out under conditions
that we expect to be the same. It is convenient to use $x_j^k$ to denote the
value of the quantity $x$ measured for system $j$ in the $k$-th assay (e.g.,
in tab.~\ref{tab:uncorrected_raw}, $x_4^2=4.78$).

The different systems can be anything, from cities to DNA sequences, from
people to chunks of metal. They can even be the same system at different times
if the quantity $x$ is expected to evolve in some reproducible manner. The
differences among the assays could be due to the experiments being performed
by the same researcher on different days, by different (but in principle
equally skilled) researchers using the same equipment, by the same researcher
using different (but in principle equally accurate) equipment, by different
(but in principle equally proficient) laboratories, etc. As long as we expect
different assays to yield the same results, their definition is compatible
with what we do here. For example, the different assays in table~II of
\citep{Galante2012}, where the production of four isoforms of
\emph{Monilophthora perniciosa} chitinase is presented, do not qualify as the
setup described here. The reason is simple: they are \emph{knowingly} carried
out at different pH and temperature. Therefore, they are naturally expected to
yield different results.

The experimental setup is thus very general, but we will introduce the
correction method as we apply it to a specific example of a \emph{real
experiment} in cell biology.

\subsection{The experiment}
\label{subsec:experiment}

The objective of the experiment is to elucidate the regulatory network of the
human protein called \emph{mitochondrial carrier homolog 1} (Mtch1), and also
\emph{presenilin 1-associated protein} (PSAP). Although this protein has been
known for almost 15 years to be involved in apoptosis \citep{Xu1999} and a
number of studies have probed its cellular function
\citep{Lamarca2008,Lamarca2007,Li2013,Mao2008,Xu2002}, not all the details are
known, specially concerning its regulation, which is uncharted territory at
the moment.

To identify binding sites for transcriptional regulators at the Mtch1 promoter
region, different DNA vectors have been constructed and transfected into
\emph{Human embryonic kidney 293T} (HEK-293T) cells. Each one of the vectors
contains a part of the Mtch1 promoter attached to a \emph{Metridia luciferase}
(\emph{MetLuc}) reporter gene. When each vector is transfected into the
HEK-293T cells, the MetLuc protein is produced and secreted to the medium,
where its activity has been measured using the Ready-to-Glow Dual Secreted
Reporter Assay Kit (Clontech). Part of this protocol involves co-transfecting
each time with a vector containing the \emph{secreted alkaline phosphatase}
(\emph{SEAP}) gene under the control of an early SV40 virus promoter. The SEAP
protein is also secreted to the medium, and the measure of its activity is
used to normalize the activity of MetLuc, with the objective of eliminating
differences in the signal due to changes in the transfection efficiency.
Hence, the activity of MetLuc is divided by that of co-transfected SEAP, and
the results are reported in \emph{relative light units} (RLU), which are the
units used in tab.~\ref{tab:uncorrected_raw} and throughout this section. The 
complete study which will be presented elsewhere.

The example we will consider here pertains only to a small part of the data
obtained for the mentioned study since it is enough for us to illustrate the
correction method. We will use the MetLuc activity values corresponding to
five vectors which contain incrementally deleted parts of the \emph{Mtch1}
promoter (denoted by pMAN12, pMAN17, pMAN18, pMAN19 and pMAN20) as well as a
control vector containing the \emph{MetLuc} gene but no promoter region at all
(pMetLuc$-$). The measured MetLuc activity values (the quantity $x$ in this
example) for the six vectors (the systems) in three assays are presented in
tab.~\ref{tab:uncorrected_raw} in sec.~\ref{sec:introduction}. This is our 
starting point.

\subsection{The problem with the results}
\label{subsec:problem}

As we advanced in sec.~\ref{sec:introduction}, the problem with the data in
tab.~\ref{tab:uncorrected_raw} begins to emerge when we compute the average of
$x$ for the system $j$ summing the results of all the assays and dividing by
the total number of assays $M$:
\begin{equation}
\label{eq:avg_sys_j}
\mu_j = \frac{1}{M} \sum_{k=1}^M x_j^k \ , \qquad j=1,2,\ldots,N \ .
\end{equation}
The corresponding standard deviation is computed as usual through:
\begin{equation}
\label{eq:stddev_sys_j}
\sigma_j = \sqrt{ \frac{1}{M} \sum_{k=1}^M \left( x_j^k - \mu_j \right)^2 }
 = \sqrt{ \frac{1}{M} \sum_{k=1}^M \left( x_j^k \right)^2
         - \left( \frac{1}{M} \sum_{k=1}^M x_j^k \right)^2}
 \ , \qquad j=1,2,\ldots,N \ .
\end{equation}
These two values are represented for all systems in the last two columns of
tab.~\ref{tab:uncorrected_raw}, and we can see there that the standard
deviations are so large that they render the results almost useless. The same
problem can be appreciated if we look in fig.~\ref{fig:errorbars_uncorrected}
(in sec.~\ref{sec:introduction}) at the bar chart associated to the last two
columns of tab.~\ref{tab:uncorrected_raw}.

\begin{table}[!b]
\begin{center}
\begin{tabular}{l|cccccc}
           & pMAN12 & pMAN17 & pMAN18 & pMAN19 & pMAN20 & pMetLuc$-$ \\
\hline
pMAN12     &    --- &  0.475 &  0.198 &  0.662 &  0.349 &      0.161 \\
pMAN17     &    --- &    --- &  0.178 &  0.399 &  0.618 &      0.121 \\
pMAN18     &    --- &    --- &    --- &  0.246 &  0.077 &      0.145 \\
pMAN19     &    --- &    --- &    --- &    --- &  0.340 &      0.215 \\
pMAN20     &    --- &    --- &    --- &    --- &    --- &      0.050 \\
pMetLuc$-$ &    --- &    --- &    --- &    --- &    --- &        ---
\end{tabular}
\caption{\label{tab:uncorrected_pvalue}{\small Probabilities (or $p$-values) 
that the observed differences between the averages $\mu_j$ and $\mu_l$ of the 
measured promoter activity ($x$ quantity) for each pair of systems (vectors) 
can be produced  by pure chance. Values smaller than 0.05 indicate that the 
observed difference is statistically significant.}}
\end{center}
\end{table}

In a more quantitative way and advancing the requirement that the inter-system
variation of $x$ is what really matters to us, we can calculate the
probability that the observed difference between two average values, $\mu_j$
and $\mu_l$, corresponding to two different vectors can be produced by pure
chance, i.e., without the need to resort to any supplementary explanation such
as the difference in the sequences of the two promoter regions in the vectors.
This probability can be obtained as the so called $p$-value associated to a
two-sample Student's $t$-test with unequal variances
\citep[p.~181]{Daniel2009}, \citep[p.~253]{Le2003}. One typically considers
the observed difference to be statistically significant when $p < 0.05$, that
is, when the probability that it can be obtained by pure chance is less than
5\% \citep{Pignatelli2003}. In tab.~\ref{tab:uncorrected_pvalue}, we present
the $p$-values associated to the activity measures of each pair of vectors in
tab.~\ref{tab:uncorrected_raw}, as computed by Microsoft Excel. We can
appreciate that our intuition about the poor quality of our results is
confirmed: Only two out of the fifteen possible pairs come close to the $p =
0.05$ threshold, none is below it, and several are significantly larger.

It is at this point when we are tempted to think that everything is lost and
just throw in the towel. Our results are bad. We have to dump them and perform
the experiments again. Period.

However, as we advanced in sec.~\ref{sec:introduction}, there are \emph{two
characteristics} about the problem we are considering here that, when
combined, can save our day.

\subsection{Requirements to apply the correction method}
\label{subsec:requirements}

The first one is related to the type of questions we are interested in making
and answering:

\begin{quote}
We are not interested in the \emph{absolute value} of $x$ for each given 
system (the MetLuc activity for each vector). What really matters to us is the 
\emph{variation} in $x$ from one system to another.
\end{quote}

For example, whether or not we could safely claim that the activity
corresponding to pMAN12 is larger than, and approximately the double of, that
associated to pMAN20. Indeed, if we \emph{are} interested in the absolute
value of MetLuc activity in RLU, the results in tab.~\ref{tab:uncorrected_raw} 
are just beyond rescue and the discussion ends here.

The second characteristic that, together with the one we just discussed, will
allow us to correct the bad looking results in tab.~\ref{tab:uncorrected_raw}
has to do with the properties of the observed measures themselves:

\begin{quote}
Even if large differences in absolute value are observed across the different
assays, the `tendency' of the variations is similarly captured in all three of
them. Technically, the different assays present \emph{high linear correlation} 
with one another.
\end{quote}

This is even more apparent in the graphical representation in
fig.~\ref{fig:tendency_uncorrected} (in sec.~\ref{sec:introduction}), and
without this kind of behavior in our data the correction method we will
introduce next would not yield satisfactory results.

\begin{figure}[!ht]
\begin{center}
\includegraphics[scale=0.25]{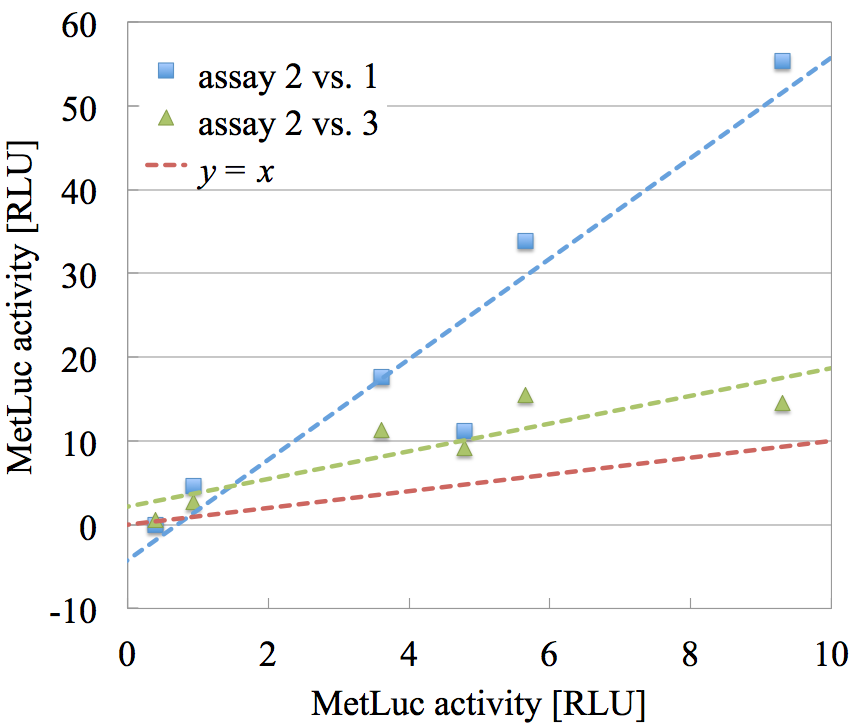}
\caption{\label{fig:fit_uncorrected}{\small Scatter plots comparing the $x$
quantity (MetLuc activity) of the six systems (vectors) in
tab.~\ref{tab:uncorrected_raw} for different pairs of assays. Using blue
squares, MetLuc activity in assay 2 vs.~the same quantity in assay 1. Using
green triangles, assay 2 vs.~assay 3. The least-squares fit lines are depicted
using the same color as the respective points, and we also show the $y = x$
line in red for reference.}}
\end{center}
\end{figure}

In fig.~\ref{fig:fit_uncorrected}, we have represented two scatter plots: both
using the values of assay 2 in the $x$-axis, one of them using the values of
assay 1 as the $y$-coordinate (blue squares), the other using the values of
assay 3 (green triangles). We have performed the two corresponding linear fits
and we have depicted the corresponding tendency lines using the same color as
the respective points. We also show the $y = x$ line in red for reference. For
the reason behind the choice of these two concrete pairs of assays, see
sec.~\ref{sec:method}.

Several points are worth remarking about this graph:

\begin{itemize}

\item As we guessed, the linear correlation between the values in the
different pairs of assays is high, with Pearson's correlation coefficient
$r=0.947$ for assays 2 vs.~1, $r=0.881$ for assays 2 vs.~3. This is the
mathematical property that embodies the intuitive property that `the different
assays similarly capture the tendency in the measured data'. Also, as we
mentioned before, this high correlation is one of the two requirements for the
method we introduce here to be applicable.

\item The fact that the fit lines have \emph{non-unit slope} is telling us 
that, although the tendency is similar across the assays, the absolute value 
is not. The two things together mean that there is a \emph{multiplicative 
systematic error} between the pairs of assays which is possible to remove. 

\item The fact that the fit lines have \emph{non-zero intercept} is telling us 
that we also have an \emph{additive multiplicative systematic error}. Our
method will eliminate it as well, as we shall see.

\end{itemize}

\subsection{The method}
\label{subsec:method}

To quantitatively assess the possibility that the data in
tab.~\ref{tab:uncorrected_raw} (or the analogous one in any experiment with
the structure described in sec.~\ref{subsec:setup}) satisfies the second
requirement in sec.~\ref{subsec:requirements}) and can therefore be corrected,
we begin by performing all least-squares linear fits between all possible
pairs of assays $k$ and $l$ [see, e.g., \citep[p.~~70]{Kirkup2006}]. For each
pair, we use the values of the first assay as the $x$ coordinate and those of
the second one as the coordinate $y$. The result of such a fit is a
\emph{tendency line} of the form:
\begin{equation}
\label{eq:best_fit_line}
y = b_{kl} x + a_{kl} \ ,
\end{equation}
where $b_{kl}$ is called the \emph{slope} and $a_{kl}$ the \emph{intercept}
(or $y$-\emph{intercept}). They are computed using the following formulas:
\begin{subequations}
\label{eq:b_and_a}
\begin{align}
b_{kl} & = \frac{\mathrm{Cov}(k,l)}{S_k^2} \ , \label{eq:b} \\
a_{kl} & = A_l - b_{kl} A_k \ , \label{eq:a}
\end{align}
\end{subequations}
where $A_k$ is the average of the measured quantity across systems and in the
one single assay $k$ [not to be confused with the averages across assays for
one single system computed using eq.~(\ref{eq:avg_sys_j}), and presented in
tab.~\ref{tab:uncorrected_raw} and fig.~\ref{fig:errorbars_uncorrected}]:
\begin{equation}
\label{eq:avg_assay_k}
A_k = \frac{1}{N} \sum_{j=1}^N x_j^k \ .
\end{equation}
Of course, $A_l$ is obtained just changing $k$ by $l$ in this expression.

The quantity $S_k$ is the standard deviation in $A_k$, given by [compare now
with eq.~(\ref{eq:stddev_sys_j})]:
\begin{equation}
\label{eq:stddev_assay_k}
S_k = \sqrt{ \frac{1}{N} \sum_{j=1}^N \left( x_j^k - A_k \right)^2 }
 = \sqrt{ \frac{1}{N} \sum_{j=1}^N \left( x_j^k \right)^2
         - \left( \frac{1}{N} \sum_{j=1}^N x_j^k \right)^2} \ ,
\end{equation}
and $\mathrm{Cov}(k,l)$ is the \emph{covariance} between the values in assay
$k$ and those in assay $l$:
\begin{equation}
\label{eq:cov_assays_kl}
\mathrm{Cov}(k,l) = \frac{1}{N} \sum_{j=1}^N
 \left(x_j^k - A_k\right) \left(x_j^l - A_l\right) \ .
\end{equation}

With these quantities in hand, we are prepared to compute the \emph{Pearson
correlation coefficient} $r_{kl}$ associated to the goodness of the linear fit
between each pair of assays $k$ and $l$, which is given by
\citep[eq.~(5.62)]{Kirkup2006}:
\begin{equation}
\label{eq:r_kl}
r_{kl} = \frac{\mathrm{Cov}(k,l)}{S_k S_l} \ .
\end{equation}

In the first three columns of tab.~\ref{tab:uncorrected_r}, we present the
Pearson correlation coefficients corresponding to each pair of assays in the
example experiment whose results can be read in 
tab.~\ref{tab:uncorrected_raw}. We can see that $r_{kl}$ is close to 1.0 for
all pairs, and we can therefore suspect that our correction method will
produce sizable improvements in the data.

\begin{table}[!ht]
\begin{center}
\begin{tabular}{l|ccc||c}
        & assay 1 & assay 2 & assay 3 & $r_k$ \\
\hline
assay 1 &   0.000 &   0.947 &   0.852 & 0.900 \\
assay 2 &     --- &    0.00 &   0.881 & 0.914 \\
assay 3 &     --- &     --- &   0.000 & 0.867
\end{tabular}
\caption{\label{tab:uncorrected_r}{\small Pearson's correlation coefficient 
$r_{lk}$ between each pair of assays in the experiment described in 
sec.~\ref{subsec:experiment}. The last column displays the average $r_k$ of 
each assay with respect to all the rest of them.}}
\end{center}
\end{table}

The first step to actually apply the method consists of selecting a
\emph{reference assay}. Since we do not know the `true' values of the $x$
quantity (MetLuc activity) for the different systems, we will compare all the
assays to the reference one and we will correct them against it.

In order to perform the selection of the reference assay with the least bias
possible, we measure `how different' each assay is to the rest and we choose
the one that is the least different; in a sense, the most representative one.
To quantify this `difference' we use in fact the Pearson correlation
coefficient, since it presents a property which makes it very convenient for
our purposes: It discounts (is insensitive to) the possible existence of both
additive and multiplicative systematic errors between the compared assays,
thus measuring the difference in the variation tendency only
\citep{Alonso2006}; which is exactly what we need. Also notice that, as a
simple consequence of its definition in eq.~(\ref{eq:r_kl}), $r_{kl}$ is
symmetric under the permutation of the indices $k$ and $l$. This is intuitive,
since it means that `the difference between assays $k$ and $l$' is the same as
`the difference between assays $l$ and $k$'.

The step that remains to be able to select the reference assay is simple: Just
compute the \emph{average correlation coefficient} $r_k$ of the $k$-th assay 
with respect to all the rest of them:
\begin{equation}
\label{eq:r_k}
r_k = \frac{1}{M - 1} \sum_{l \neq k} r_{kl} \ ,
\end{equation}
and pick the one with the \emph{largest} $r_k$.

In the last column of tab.~\ref{tab:uncorrected_r}, we show the average
correlation coefficient $r_k$ associated to each assay. We can see that it is
the largest for assay 2. Therefore, we select assay 2 as our reference assay
in the example we are discussing (which, by the way, explains the particular
fits portrayed in fig.~\ref{fig:fit_uncorrected}).

Now that the reference assay has been chosen and all the linear fits have been
computed, we are ready to apply the \emph{correction} to the rest of assays.
If we denote by $f$ the value of the index $k$ that corresponds to the 
reference assay ($f = 2$ in our example) and we use $\tilde{x}_j^l$ for the 
corrected value associated to the original quantity $x_j^l$ (system $j$, assay 
$l$), the \emph{correction formula} reads like this:
\begin{equation}
\label{eq:main}
\tilde{x}_j^l = \frac{x_j^l - A_l}{b_{fl}} + A_f \ .
\end{equation}
In order to produce the whole set of corrected results, we should apply this
for all assays $l \neq f$, with $l=1,\ldots,M$, and for all systems with the
index $j=1,\ldots,N$.

In order to understand the reason behind this formula, it is convenient to
write the inverse transformation by solving for $x_j^l$:
\begin{equation}
\label{eq:inverse}
x_j^l = b_{fl} \big(\tilde{x}_j^l - A_f\big) + A_l \ ,
\end{equation}
and also to notice that the systems-average of $\tilde{x}_j^l$ is given by:
\begin{equation}
\label{eq:avg_assay_l_z}
\tilde{A}_l = \frac{1}{N} \sum_{j=1}^N \tilde{x}_j^l
 = \frac{(1/N)\sum_{j=1}^N x_j^l - A_l}{b_{fl}} + A_f
 = \frac{A_l - A_l}{b_{fl}} + A_f = A_f \ ,
\end{equation}
i.e., all the averages of the corrected assays are equal to the average of the
reference one. Now, if we take eq.~(\ref{eq:inverse}) to the covariance in
eq.~(\ref{eq:cov_assays_kl}) with $k=f$, we obtain:
\begin{eqnarray}
\label{eq:cov_assays_fl}
\mathrm{Cov}(f,l) & = & \frac{1}{N} \sum_{j=1}^N
 \left(x_j^f - A_f\right) \left(x_j^l - A_l\right)
 = \frac{1}{N} \sum_{j=1}^N \left(x_j^f - A_f\right)
   \Big(b_{fl} \big[\tilde{x}_j^l - A_f\big] + A_l - A_l\Big)
 \nonumber \\
 & = & b_{fl} \frac{1}{N} \sum_{j=1}^N \left(x_j^f - A_f\right)
                          \left(\tilde{x}_j^l - A_f\right)
 =  b_{fl} \frac{1}{N} \sum_{j=1}^N \left(\tilde{x}_j^f - \tilde{A}_f\right)
                       \left(\tilde{x}_j^l - \tilde{A}_l\right)
 \nonumber \\
 & = & b_{fl} \mathrm{Cov}(\tilde{f},\tilde{l}) \ ,
\end{eqnarray}
where, in the last step of the second line, we have used that $A_f =
\tilde{A}_l$ [as we proved in eq.~(\ref{eq:avg_assay_l_z})], but also that the
correction in eq.~(\ref{eq:main}) is obviously the identity for the reference
assay $f$ (it suffices to notice that $b_{ff}=1$), which makes $x_j^f =
\tilde{x}_j^f$, as well as all the derived quantities, such as $A_f =
\tilde{A}_f$. In the last line of eq.~(\ref{eq:cov_assays_fl}), we have simply
used the natural notation $\mathrm{Cov}(\tilde{f},\tilde{l})$ to indicate the
covariance between the \emph{corrected} assays $\tilde{f}$ and $\tilde{l}$.
Finally, if we use eq.~(\ref{eq:cov_assays_fl}) together with the definition 
of the slope in eq.~(\ref{eq:b}) (with $k=f$), we obtain:
\begin{equation}
\label{eq:b_corrected}
\tilde{b}_{fl} = \frac{\mathrm{Cov}(\tilde{f},\tilde{l})}{\tilde{S}_f^2}
 = \frac{1}{b_{fl}} \frac{\mathrm{Cov}(f,l)}{S_f^2}
 = \frac{b_{fl}}{b_{fl}} = 1 \ ,
\end{equation}
where we have denoted by $\tilde{b}_{fl}$ the slope associated to the fit
between the corrected assays $\tilde{f}$ and $\tilde{l}$, and we have used 
that $\tilde{S}_f = S_f$. Also, it is easy to prove that:
\begin{equation}
\label{eq:a_corrected}
\tilde{a}_{fl} = \tilde{A}_l - \tilde{b}_{fl} \tilde{A}_f
 = \tilde{A}_l - \tilde{A}_f
 = \tilde{A}_f - \tilde{A}_f = 0 \ .
\end{equation}

That is, the slope of the fits among the corrected assays is 1 and the
intercept is 0. Since we argued that the first can be interpreted as a
multiplicative systematic error and the second as an additive one, we have
just proved that our proposed correction in eq.~(\ref{eq:main}) has the
promised effect of eliminating both errors. To see that this also has the
effect of reducing the standard deviations and improving the statistical
significance of our results, we turn to the next section.

But before, let us mention a final consistency property of the correction
method: In mathematical jargon, it is \emph{idempotent}. In plain words,
applying it twice is the same as applying it once, i.e., if we apply the
whole correction process to the corrected results, we find that nothing 
changes. The corrected-corrected results are just the corrected results.

All the formulae needed to compute the linear fits, the inter-assay
correlation coefficients, as well as the correction in eq.~(\ref{eq:main}) are
provided in this section and they are very simple. The reader can choose to
implement them in any spreadsheet of her liking, or she can use the Perl
scripts we have written for the occasion and which can be found in the
\href{https://drive.google.com/folderview?id=0B0fpdP1w5u6bYW0xU0lnYzB4UWs&usp=sharing}{supplementary material}. Also in the \href{https://drive.google.com/folderview?id=0B0fpdP1w5u6bYW0xU0lnYzB4UWs&usp=sharing}{supplementary material}, we provide a cheat sheet with the bare steps 
of our method, conveniently organized, briefly stated, and stripped off of all 
the explanatory text that surrounds the steps in this article.

\section{Results}
\label{sec:results}

If we apply the correction in eq.~(\ref{eq:main}) to our original results in
tab.~\ref{tab:uncorrected_raw}, we find the corrected values in the second
table of tab.~\ref{tab:before_after_raw} (where we have repeated the 
uncorrected data to facilitate the comparison).

\begin{table}[!ht]
\begin{center}
\textbf{Before}\\[5pt]
\begin{tabular}{l||rrr|r@{\hspace{3pt}}c@{\hspace{3pt}}r}
           & assay 1 & assay 2 & assay 3 & $\mu$ & $\pm$ & $\sigma$ \\
\hline
pMAN12     &   33.88 &    5.65 &   15.53 & 18.36 & $\pm$ &    14.33 \\
pMAN17     &   17.60 &    3.61 &   11.29 & 10.83 & $\pm$ &     7.01 \\
pMAN18     &    4.62 &    0.94 &    2.72 &  2.76 & $\pm$ &     1.84 \\
pMAN19     &   55.35 &    9.30 &   14.52 & 26.39 & $\pm$ &    25.22 \\
pMAN20     &   11.15 &    4.78 &    9.10 &  8.35 & $\pm$ &     3.52 \\
pMetLuc$-$ &    0.00 &    0.39 &    0.54 &  0.31 & $\pm$ &     0.28
\end{tabular}\\[10pt]
\textbf{After}\\[5pt]
\begin{tabular}{l||rrr|r@{\hspace{3pt}}c@{\hspace{3pt}}r}
           & assay 1 & assay 2 & assay 3 & $\tilde{\mu}$ & $\pm$ & $\tilde{\sigma}$ \\
\hline
pMAN12     &    6.35 &    5.65 &    8.10 &  6.70 & $\pm$ &     1.26 \\
pMAN17     &    3.64 &    3.61 &    5.52 &  4.26 & $\pm$ &     1.10 \\
pMAN18     &    1.48 &    0.94 &    0.34 &  0.92 & $\pm$ &     0.57 \\
pMAN19     &    9.93 &    9.30 &    7.48 &  8.91 & $\pm$ &     1.27 \\
pMAN20     &    2.56 &    4.78 &    4.20 &  3.85 & $\pm$ &     1.15 \\
pMetLuc$-$ &    0.71 &    0.39 & $-$0.98 &  0.04 & $\pm$ &     0.90
\end{tabular}
\caption{\label{tab:before_after_raw}{\small Activity of the MetLuc protein
under the control of six different promoter sequences measured in three
assays, before and after the correction described in sec.~\ref{subsec:method}. The last two columns correspond to the average $\mu$ of the three
assays for each vector, and the associated standard deviation (or error)
$\sigma$. The units as well as the rest of the experiment's details are
described in the text.}}
\end{center}
\end{table}

\begin{figure}[!ht]
\begin{center}
\includegraphics[scale=0.3]{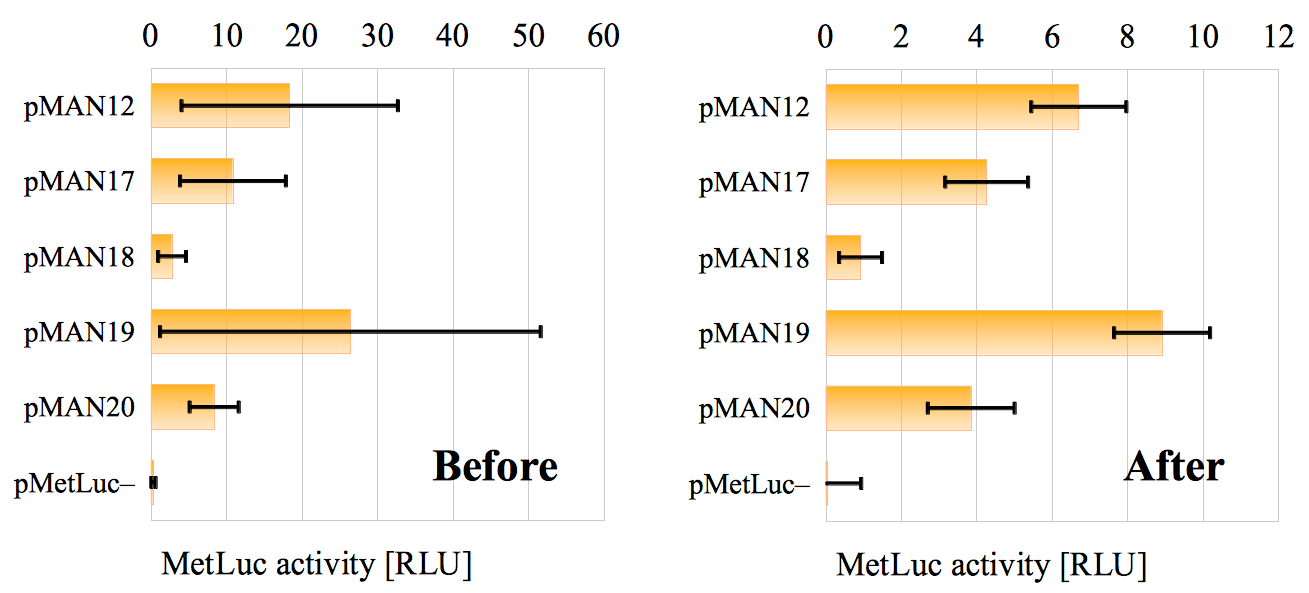}
\caption{\label{fig:errorbars_before_after}{\small Bar chart representation of 
the average values $\mu$ (orange bars) and the associated standard deviation
$\sigma$ (black capped lines) in tab.~\ref{tab:before_after_raw}, before and 
after the correction described in sec.~\ref{subsec:method}.}}
\end{center}
\end{figure}

At first sight, the corrected standard deviations $\tilde{\sigma}$ seem much 
better when compared to their associated averages $\tilde{\mu}$ for each 
system. This impression is reinforced if we take a look at the corresponding
bar charts in fig.~\ref{fig:errorbars_before_after}.

If we want to be more quantitative, and recalling that the inter-system
variation of MetLuc activity is what really matters to us, we can repeat the
$p$-values calculation in sec.~\ref{subsec:problem}, this time for the
corrected data. In tab.~\ref{tab:before_after_pvalue}, we present both the
original $p$-values obtained from the uncorrected results as well as the new
ones. We remind the reader that the $p$-value's meaning is that it quantifies
the probability that the observed difference between two average values,
$\mu_j$ and $\mu_l$, corresponding to two different vectors can be produced by
pure chance, i.e., without the need to resort to any supplementary explanation
such as the difference in the sequences of the two promoter regions in the
vectors. One typically considers the observed difference to be statistically
significant when $p < 0.05$, that is, when the probability that it can be
obtained by pure chance is less than 5\%. As we can see in
tab.~\ref{tab:before_after_pvalue}, while the original situation was
despairing, with two out of the fifteen possible pairs close to the $p = 0.05$
threshold, none below it, and several significantly larger, the corrected
$p$-values show a much better behavior. For the corrected data, eleven out of
the fifteen possible comparisons are below the $p = 0.05$ threshold, two of
them are close to it, and only two are significantly larger. This means that
most of the observed differences in MetLuc activity are now statistically
significant.

\begin{table}[!ht]
\begin{center}
\textbf{Before}\\[5pt]
\begin{tabular}{l|cccccc}
           & pMAN12 & pMAN17 & pMAN18 & pMAN19 & pMAN20 & pMetLuc$-$ \\
\hline
pMAN12     &    --- &  0.475 &  0.198 &  0.662 &  0.349 &      0.161 \\
pMAN17     &    --- &    --- &  0.178 &  0.399 &  0.618 &      0.121 \\
pMAN18     &    --- &    --- &    --- &  0.246 &  0.077 &      0.145 \\
pMAN19     &    --- &    --- &    --- &    --- &  0.340 &      0.215 \\
pMAN20     &    --- &    --- &    --- &    --- &    --- &      0.050 \\
pMetLuc$-$ &    --- &    --- &    --- &    --- &    --- &        ---
\end{tabular}\\[10pt]
\textbf{After}\\[5pt]
\begin{tabular}{l|cccccc}
           & pMAN12 & pMAN17 & pMAN18 & pMAN19 & pMAN20 & pMetLuc$-$ \\
\hline
pMAN12     &    --- &  0.066 &  \textbf{0.007} &  0.100 &  \textbf{0.045} &      \textbf{0.003} \\
pMAN17     &    --- &    --- &  \textbf{0.018} &  \textbf{0.009} &  0.679 &      \textbf{0.007} \\
pMAN18     &    --- &    --- &    --- &  \textbf{0.003} &  \textbf{0.030} &      0.236 \\
pMAN19     &    --- &    --- &    --- &    --- &  \textbf{0.007} &      \textbf{0.001} \\
pMAN20     &    --- &    --- &    --- &    --- &    --- &      \textbf{0.012} \\
pMetLuc$-$ &    --- &    --- &    --- &    --- &    --- &        ---
\end{tabular}
\caption{\label{tab:before_after_pvalue}{\small Probabilities (or $p$-values)
that the observed differences between the averages $\mu_j$ and $\mu_l$ of the
measured promoter activity for each pair of vectors can be produced by pure
chance. The two tables correspond to the data before and after the correction
described in sec.~\ref{subsec:method}. Values smaller than 0.05 indicate that
the observed difference is statistically significant in both cases, and the
entries satisfying this condition have been highlighted using boldface 
fonts.}}
\end{center}
\end{table}

\begin{figure}[!t]
\begin{center}
\includegraphics[scale=0.27]{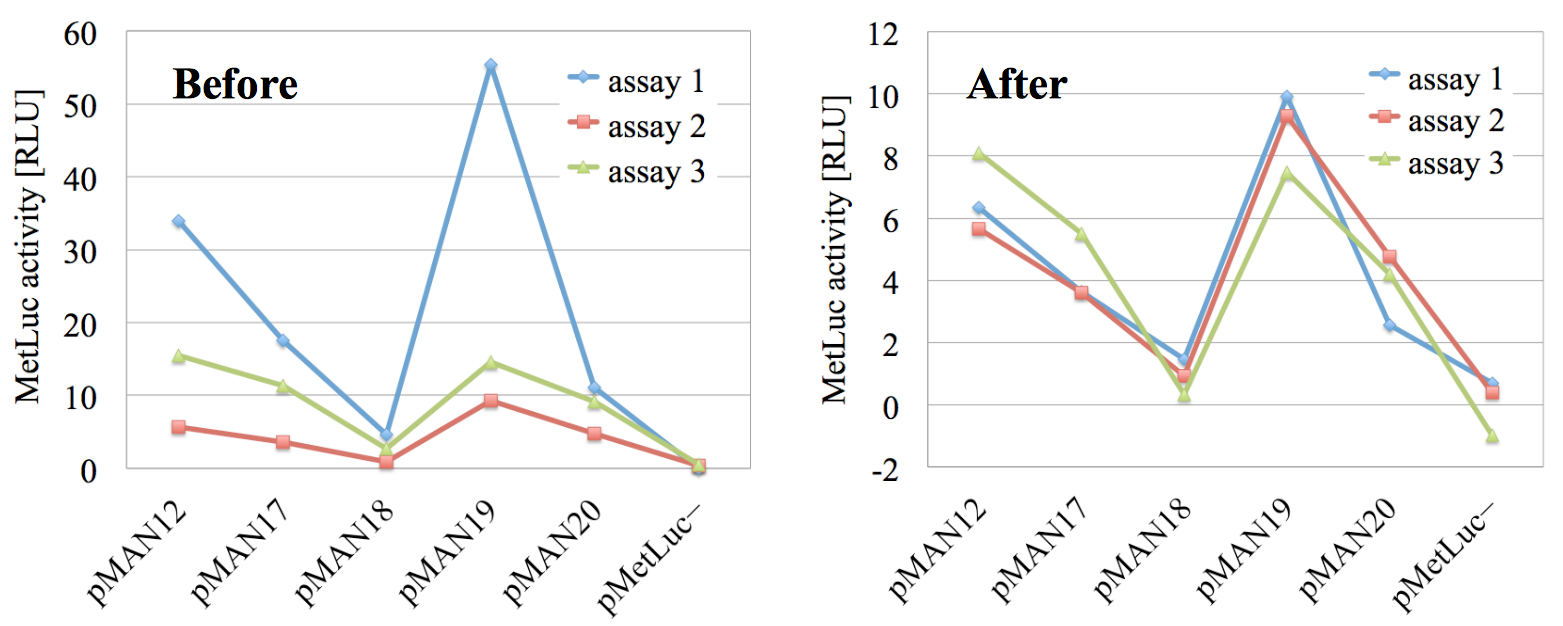}
\caption{\label{fig:tendency_before_after}{\small Variation of the quantity
$x$ (MetLuc activity) in tab.~\ref{tab:uncorrected_raw} for the six systems
(vectors) studied, before and after the correction described in
sec.~\ref{subsec:method}. Each color corresponds to a different assay, and the
lines joining the experimental points have been added for visual comfort.}}
\end{center}
\end{figure}

In order to enrich our picture of what is going on here, we can also take a
look at the corrected version of the tendency plot that we presented before in
fig.~\ref{fig:tendency_uncorrected} and which we now repeat here on the left
of fig.~\ref{fig:tendency_before_after}. As we can see in the corrected
tendency plot on the right, the fact that all three assays correctly captured
the overall variation tendency of the data has been maximally leveraged by the
correction in eq.~(\ref{eq:main}). Without altering the legitimate random
noise in the original results, the additive and multiplicative systematic
errors have been eliminated, and the corrected tendency lines are now 
optimally superimposed.

\begin{figure}[!ht]
\begin{center}
\includegraphics[scale=0.27]{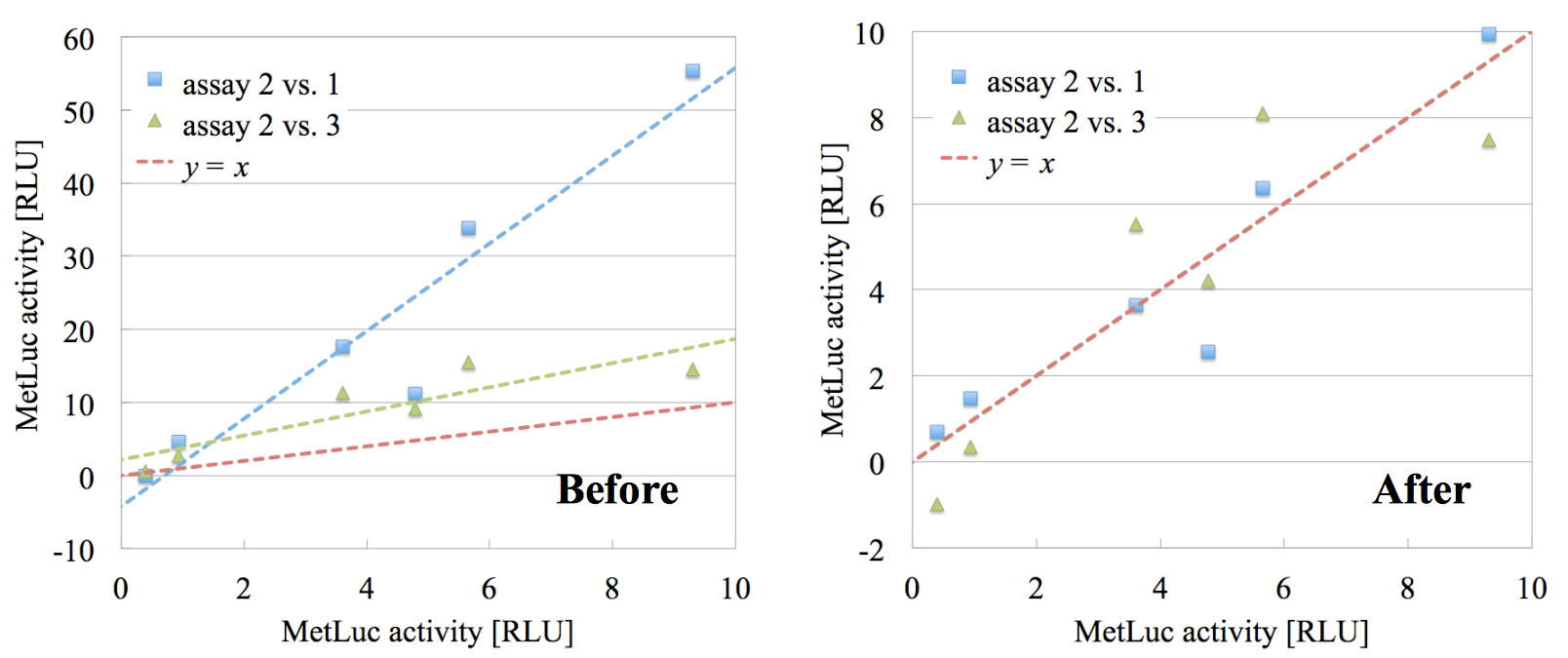}
\caption{\label{fig:fit_before_after}{\small Scatter plots comparing the $x$
quantity (MetLuc activity) of the six systems (vectors) in
tab.~\ref{tab:before_after_raw} for different pairs of assays, before and
after the correction described in sec.~\ref{subsec:method}. Using blue
squares: MetLuc activity in assay 2 vs.~the same quantity in assay 1. Using
green triangles: assay 2 vs.~assay 3. The least-squares fit lines are depicted
using the same color as the respective points, and we also show the $y = x$
line in red for reference.}}
\end{center}
\end{figure}

Similarly, we can compare the original and corrected scatter plots in
fig.~\ref{fig:fit_before_after}. In the second one, the best fit lines
corresponding to assays 2 vs.~1 and assays 2 vs.~3 have been omitted because
they coincide with the zero-intercept unit-slope $y=x$ line. This is the
precise mathematical embodiment of the fact that the correction in
eq.~(\ref{eq:main}) `eliminates the additive and multiplicative errors': it
transforms all the fits against the reference assay from non-zero intercept
and non-unit slope to zero intercept and unit slope. The fact that the random
error is unmodified can be appreciated by the remaining dispersion of the
scatter plot points with respect to the $y=x$ line in the second graph in
fig.~\ref{fig:fit_before_after}.

\section{Discussion}
\label{sec:discussion}

We have just introduced a simple method for correcting the results of
multi-assay experiments which, under two very basic conditions (that only
inter-system variations matter to us, and that the different assays present
high linear correlation with one another), allows us to considerably reduce
the standard deviation of the systems' averages across assays, consequently
increasing the statistical significance of the results. We have applied the
correction method to a real experiment in cell biology where we have
appreciated a great improvement.

Our interpretation of the situation is as follows: Uncontrolled differences
(errors) appear when a given experiment is repeated. Some of them are random
(i.e., we see no pattern in them) and cannot be eliminated. Some others are
systematic and can be. If we represent a scatter plot in which the results of
one assay are placed on the $x$-axis, the results of a different one are
placed on the $y$-axis, and we perform a linear fit, we can expect to observe
two different situations:

\begin{itemize}

\item The best fit line has zero $y$-intercept and unit slope. We interpret
this as all the error being random, and no correcting action can be taken 
here. The data must be used `as is'.

\item The best fit line has non-zero intercept or non-unit slope or both. We
interpret this as some of the error being systematic, some of it random. The
non-zero intercept signals an additive systematic error; the non-unit slope a
multiplicative systematic one; and the dispersion of the scatter plot points
from the fit line signals the part of the error that is random. In such a 
case, we can apply the correction in eq.~(\ref{eq:main}), thus eliminating
both systematic components and reducing the situation to the one described in
the previous point.

\end{itemize}

As it is always the case with systematic errors, one might or might not know
the actual reasons behind them (we left the apparatus on too much time, the
cell number was larger than usual, we inadvertently used the wrong pipette,
etc.), but we do not really need to know the reasons to confidently assert
that a systematic error is indeed there. If the difference between two assays
is (mostly) captured by multiplying the results of one of them by a number $b
\neq 1$ and adding a number $a \neq 0$, we are entitled to entertain the
strong suspicion that some very real causes are behind this predictable
pattern. Hence, even if we do not know these causes, it would be a wasted
opportunity not to apply the correction in eq.~(\ref{eq:main}). If you do know
the causes, good for you. So much the better. In fact, by applying the
reasoning associated to the method described here, the presence of a non-zero
intercept or a non-unit slope in the fits of the different pairs of assays
(plus a high linear correlation among them) may suggest to the experimenter
that some additive or multiplicative systematic error is being made from assay
to assay. With this clue, she can then proceed to look for the actual
experimental causes behind them (in the case that they were previously 
unknown).

Also notice that systematic errors might not end at the linear order. The
relation between the results of two different assays could be well described
for example by a quadratic relation $y = cx^2 + bx + a$ plus some random
error; or even by higher order polynomials. No \emph{a priori} reason can
reject this possibility, however, a treatment of these more complicated cases
is outside the scope of this work.

An important part of the method introduced here is that, since we do not know
the `real' absolute value of the measured quantities (and in fact it does not
matter to us), we have to choose a \emph{reference assay} to fit all the rest
of assays to. The most reasonable way to perform this choice in an unbiased
manner is to select the most representative assay in the experiment, the one
that is `most similar to all the others'. We make this condition precise by
measuring the difference of every assay to all the rest of them and choosing
the one that is the least different to all the others. To this end, we use the
Pearson correlation coefficient associated to the goodness of the linear fit
because it correctly discounts the additive and multiplicative systematic
errors.

Also, it is worth mentioning that the use of the word `error' for the
differences between one specific assay and the rest of them might seem
unorthodox at first sight. After all, the `error' is ideally defined as the
difference between the measured quantities and their `real' values. However,
we think that this apparent overuse of the term is just that: apparent. Since
the `real' values are never actually known, the ideal definition of `error' is
philosophically appealing but practically inapplicable. What researchers
\emph{always} do is to compare one set of measures to some more accurate ones
(but not `real' yet), to some theoretical prediction (not `real' either), etc.
In this sense, and given that the `real' values of the quantity $x$ are
unknown in our experimental setup in sec.~\ref{subsec:setup} (as in all
setups!), the `best' guess we a priori have (before the proposed correction)
of the most accurate set of measures is precisely the most representative of
our assays, i.e., the one that is the least different from the rest. This is
why we choose it as the reference to which all the rest of the assays are
compared, and this is why the observed differences deserve to be intuitively
called `errors'.

Although all this seems quite straightforward, we have only found in the
literature one related proposal for a correction method that could be compared
to the one we introduce in this work (even if the rationale is never clearly
expressed as we do here). This related method readily comes to mind and it
consists of dividing, in each assay, the value of $x$ for all systems by the
value of one of them. For example, we could select pMAN18 as our
\emph{normalizing vector}, divide the activities of all the vectors in each
assay by the activity of pMAN18 in the same assay, and thus obtain a new set
of results now expressed as a \emph{normalized fold change in activity} with
respect to the pMAN18 value (which now becomes 1.0). This is used for example
in \citep{Schagat2007,YsebrantdeLendonck2013,Matsunoshita2011,Alvarez2012} or \citep[fig.~S3]{Zhang2012}.

The result of applying this normalization to the original data in
tab.~\ref{tab:normalized_raw} is presented in tab.~\ref{tab:normalized_raw}.
We see that the standard deviations have been reduced and in fact the overall
improvement is similar to what we obtained when applying the correction method
introduced in this work. However, this normalization procedure presents some
drawbacks that, in our opinion, render it inferior to our method. Namely:

\begin{table}[!t]
\begin{center}
\begin{tabular}{l||rrr|r@{\hspace{3pt}}c@{\hspace{3pt}}r}
           & assay 1 & assay 2 & assay 3 & $\mu$ & $\pm$ & $\sigma$ \\
\hline
pMAN12     &    7.33 &    6.01 &    5.71 &  6.35 & $\pm$ &     0.86 \\
pMAN17     &    3.81 &    3.84 &    4.15 &  3.93 & $\pm$ &     0.19 \\
pMAN18     &    1.00 &    1.00 &    1.00 &  1.00 & $\pm$ &     0.00 \\
pMAN19     &   11.98 &    9.89 &    5.34 &  9.07 & $\pm$ &     3.40 \\
pMAN20     &    2.41 &    5.09 &    3.35 &  3.62 & $\pm$ &     1.36 \\
pMetLuc$-$ &    0.00 &    0.41 &    0.20 &  0.20 & $\pm$ &     0.21
\end{tabular}
\caption{\label{tab:normalized_raw}{\small Fold change in activity of the
MetLuc protein under the control of six different promoter sequences measured
in three assays. The numbers in this table have been obtained from the
activity data in tab.~\ref{tab:uncorrected_raw} through division by the value
for pMAN18.}}
\end{center}
\end{table}

\begin{itemize}

\item It demands an arbitrary choice (that of the normalizing system) which
seems ad hoc and prevents automatization in some degree. Related to this, the
fact that the corrected result for the normalizing system has zero standard 
deviation does not seem easy to interpret, nor completely legitimate.

\item If we recall the general formula for the propagation of errors
\citep[p.~50]{Kirkup2006},
\begin{equation}
\label{eq:propagation_errors_gen}
s_f^2 = \sum_{i=1}^n
 \left( \frac{\partial f}{\partial q_i} \right)^2 s_i^2 \ ,
\end{equation}
where $f(q_1,\ldots,q_n)$ is a function of $n$ random variables with standard
deviations (errors) $s_1,\ldots,s_n$, we can use it to compute the error in
the normalized quantity $y_j = x_j/x^*$, where $x_j$ is the measured result
for the system $j$ (in a given assay) and $x^*$ is the quantity measured for
the system chosen to normalize the results:
\begin{equation}
\label{eq:propagation_errors_normalized}
s_{y_j}^2 = \frac{1}{(x^*)^2} s_{x_j}^2 + \frac{x_j^2}{(x^*)^4} s_{x^*}^2
 = y_j^2 \left( \frac{s_{x_j}^2}{x_j^2} + \frac{s_{x^*}^2}{(x^*)^2} \right)
 \ \Rightarrow \
\frac{s_{y_j}^2}{y_j^2}
 = \frac{s_{x_j}^2}{x_j^2} + \frac{s_{x^*}^2}{(x^*)^2} \ .
\end{equation}
We see that the error in the normalized quantity $y_j$ relative to the value
of $y_j$ itself is the sum of the relative errors of $x_j$ and $x^*$. Now, if
we happen to choose a particular normalizing system with high relative error,
this could spoil the whole assay when we divide all the results by $x^*$, even
if the rest of measures were accurate.

\item The described normalizing procedure seems fit to eliminate 
multiplicative systematic errors, but not additive ones.

\end{itemize}

Our method suffers from none of these problems:

\begin{itemize}

\item No choice of a `special' normalizing system is needed. (There is a
choice of a reference assay, but it is made in a justified way, as we have
explained.)

\item In a manner of speaking, it distributes the normalization among all the 
values in a given assay, thus minimizing the probability that one specially 
bad apple spoils the whole basket.

\item It eliminates both multiplicative and additive systematic errors.

\end{itemize}

If we check exhaustive textbooks in biostatistics, such as
\citep{Daniel2009,Vittinghoff2005,Le2003}, or more wide ranging ones, such as
\citep{Walpole2012,Kutner2005,Mickey2004,Taylor1997,Mandel1984}, we do not 
find any account of a correcting method that is similar to what we propose 
here. Some of the texts come close sometimes, but they never hit the target.

One way in which they often come close is when they discuss \emph{repeated
measures}. See for example \citep[chap.~9]{Mickey2004},
\citep[chap.~27]{Kutner2005}, or \citep{Cnaan1997},
\citep[chap.~8]{Vittinghoff2005}, and \citep[p.~346]{Daniel2009} for detailed
discussions of the concept in biosciences. `Repeated measures' consists of an
experimental setup very similar to the one used here and described in
sec.~\ref{subsec:setup}, i.e., measuring the same quantity on $N$ systems and
repeating the experiment $M$ times, but it contains a fundamental difference:
it tackles measurements \emph{that are expected to change from repetition to
repetition} [e.g., a time series, or table~II of \citep{Galante2012} discussed
in sec.~\ref{subsec:setup}]. It is a key of our setup that we expect the
results of several repetitions to be \emph{the same}. This is why it makes
sense for us to correct them, which would be unnatural in the
repeated-measures setup. Also, for repeated measures, it is not a requirement
that we are not interested in the absolute value but only in the inter-system
variation. In our case, this is essential.

In \citep[p.~539]{Walpole2012}, another similar situation to the one we have
considered here is dealt with, namely \emph{blocking}, however, they do not
discuss what to do if there is an obvious linear correlation between the
blocks (as in their figure~13.6a). Their example in figure~13.12 also seems
ripe to apply our method, but they take no correcting action on it.

One of the reasons that we imagine could be behind the fact that no precedents
of our straightforward method are found in the literature (as far as we have
been able to scan it) has to do with the usual interpretation of the range of
application of the least-squares fit protocol. Typically, fitting some values
in the $x$-axis against those on the $y$-axis is used to assess a possible
linear relationship between \emph{two different quantities} (apples and
oranges, say). So much so that $x$ is typically called the \emph{independent
variable}, while $y$ is the \emph{dependent} one. In our approach, it is a key
conceptual step to realize that it actually makes sense to investigate the
linear correlation of some quantity \emph{with itself} (measured in two
different assays), and consequently interpret any difference between the two
as experimental error (in the manner we explained before).

Another reason that is possibly behind the absence of precedents is the fact
that, despite being quite intuitive to us, systematic errors of the
\emph{multiplicative} kind are very rarely discussed in the literature.
Systematic errors are normally considered to be additive.

After a thorough search we have only found anecdotal mentions in a paper that
discusses the influence of natural fires on the air pollution of the Moscow
area \citep{Konovalov2011}, in a proceedings paper about anticorrosion coating
\citep{Niedostatkiewicz2006}, in a recent work concerned with calibration of
spectrographs for detecting earth-mass planets around sun-like stars
\citep{Glenday2012}, and in a similar paper focused in the detection and study
of quasars \citep{Johansson2000}. In all these works the authors consider the
possibility of a multiplicative systematic error in their models or
measurements, but they take no action to correct it.

Something very similar happens in \citep[p.~3]{Meloun1993}, where the
existence of multiplicative systematic errors is acknowledged in the context
of analytical chemistry, as well as the necessity to eliminate them. In
\citep{Doerffel1994}, the possibility of both additive and multiplicative
systematic errors is discussed, as well as their respective relation with
non-zero $y$-intercepts and non-unit slopes. Finally, in
\citep[p.~39]{Kirkup2006}, the authors not only discuss multiplicative
systematic errors (which they also call \emph{gain shifts} or \emph{gain
errors}), but they provide several examples where this multiplicative
systematic error can appear. Although more space is dedicated in these last
three works to the discussion of multiplicative systematic errors, the authors
do not provide any method for eliminating them either.

In addition, it is worth mentioning that, in \citep{Doerffel1994} and in
\citep{Kirkup2006}, the authors consider the error to be defined with respect
to `true' (or at least more accurate) results; in the first case to calibrate
experimental protocols, in the second one to calibrate measuring devices. As
we explained when discussing the choice of the reference assay, our 
perspective on this issue is different, and so it is the approach. For 
example, if you want to correct your results against some `better' data, you
are presumably interested not only in the variations of the measured quantity,
but also in its absolute value.

We have only found one work, concerned with gas electron diffraction data
\citep{Gundersen1998}, in which the authors \emph{both} consider the existence
of multiplicative systematic errors \emph{and} take actions to correct them.
However, the proposed correction is particular to the concrete problem
studied, and the experimental setup is different to the one described in
sec.~\ref{subsec:setup}: The authors refer to systematic errors in
experimental data with respect to the `true' values, not to systematic errors
between different measures of the same quantity as we do here.

\section{Conclusions}
\label{sec:conclusions}

We have introduced a method for correcting the data in experiments in which a
single quantity $x$ is measured for a number of systems in multiple
repetitions or assays. If we are not interested in the absolute value of $x$
but only in the inter-system variations, and the results in different assays
are highly correlated with one another, we can use the proposed method to
eliminate both additive and systematic differences (errors) between each one
of the assays and a suitably chosen reference one. As we have shown using a
real example of a cell biology experiment, this correction can
considerably reduce the standard deviation in the systems' averages across
assays, and consequently improve the statistical significance of the data.

The method is of very general applicability, not only to experimental results
but possibly also to numerical simulations, as long as the structure of the
setup and the requirements on the data are those just mentioned and carefully
discussed in sec.~\ref{subsec:setup}. This, together with its simplicity of
application (the only mathematical infrastructure needed to apply it is
basically least-squares linear fits), makes the method of very wide interest
in any quantitative scientific field that deals with data subject to
uncertainty.

Some possible lines of future work include the application of the method to a
wider variety of problems, a deeper statistical analysis of its properties and
the assumptions behind it, or the extension to systematic differences of
higher-than-linear order that we briefly mentioned in
sec.~\ref{sec:discussion}.

\section*{Acknowledgements}

\hspace{0.5cm} We would like to thank Professors Jes\'us Pe\~na, Silvano Pino,
Juan Puig, Ricardo Rosales and Javier Sancho for recommending to us the
reference statistics and biostatistics textbooks that we have used in the 
writing of the manuscript.

This work has been supported by the grants FIS2009-13364-C02-01 (Ministerio de
Ciencia e Innovaci\'on, Spain), UZ2012-CIE-06 (Universidad de Zaragoza,
Spain), Grupo Consolidado ``Biocomputaci\'on y F\'{\i}sica de Sistemas
Complejos'' (DGA, Spain), also by grants BFU2009-11800 (Ministerio de Ciencia
e Innovaci\'on, Spain), and UZ2010-BIO-03 and UZ2011-BIO-02 (Universidad de
Zaragoza, Spain) to J.A.C.


\end{document}